\documentclass[sigconf]{acmart}

\usepackage{xspace}
\usepackage[ruled,vlined]{algorithm2e}
\usepackage{array}
\usepackage{xcolor}
\usepackage{optidef}
\usepackage{subfigure}
\usepackage{framed} 
\usepackage{wrapfig}
\newcolumntype{L}{>{$}l<{$}} 
\newcolumntype{R}{>{$}r<{$}} 
\usepackage{enumitem}

\AtBeginDocument{%
  \providecommand\BibTeX{{%
    \normalfont B\kern-0.5em{\scshape i\kern-0.25em b}\kern-0.8em\TeX}}}

\setcopyright{rightsretained}
\acmDOI{}
\acmISBN{}

\acmConference[]{arXiv preprint}{June}{2020}

\begin{document}

\title{\system: Towards a Generic, Fully-Automated Test and Validation Platform for Wireless Networks}


\author{Kerim G\"okarslan}
\affiliation{%
  \institution{Department of Computer Science, Yale University}
  \city{New Haven}
  \state{Connecticut}
}
\email{kerim.gokarslan@aya.yale.edu}

\renewcommand{\shortauthors}{Kerim G\"okarslan}
\newcommand{\kerim}[1]{\textcolor{blue}{#1}}
\newcommand{\eay}[1]{\textcolor{red}{#1}}
\newcommand{\system}{\emph{Menes}\xspace}
\newcommand{\para}[1]{\noindent {\bf #1}}
\begin{abstract}

A major step in developing robust wireless systems is to test and validate the design under a variety of circumstances. As wireless networks become more complex, it is impractical to perform testing on a real deployment. As a result, the network administrators rely on network simulators or network emulators to validate their configurations and design. Unfortunately, network simulation falls short per it requires users to model the network behavior analytically. On the other hand, network emulation allows users to employ real network applications on virtualized network devices. Despite their complex design, the existing network emulation solutions miss full-scale automation rather they rely on experienced users to write complex configuration scripts making testing. Therefore, the validation process is prone to human operator errors. Furthermore, they require a significant amount of computational resources that might not be feasible for many users. Moreover, most network emulators focus on lower layers of the network thus requiring users to employ their own network applications to control and measure network performance. To overcome these challenges, we propose a novel wireless network emulation platform, the system, that provides users a unified, high-level configuration interface for different layers of wireless networks to reduce management complexities of network emulators while having a generic, fully-automated platform. Menes is a generic, full-stack, fully-automated test and validation platform that empowers existing state-of-the-art emulation, virtualization, and network applications including performance measurement tools. We then provide an implementation of Menes based on the Extendable Mobile Ad-hoc Network Emulator (EMANE) with Docker containers. Our extensive evaluations show that the system requires much less computing resources, significantly decreases capital expenses (CAPEX) and operational expenses (OPEX), and greatly extensible for different use cases.

\end{abstract}

\settopmatter{printacmref=false, printccs=false, printfolios=true}
\maketitle
\section{Introduction}


The use of wireless communication technologies has been increasingly popular since its first introduction with the analog radio over two centuries ago. Wireless technologies have been heavily used by the military by the end of the 1980s except for the radio and television broadcasting~\cite{rappaport1991wireless,seymour2011history}. With the introduction of cellular networks for citizens and development of wireless personal area networking (WPAN) technologies such as IEEE 802.11~\cite{10.5555/553211} and Bluetooth~\cite{bluetooth}, complex and heterogeneous wireless networks have been evolved. Apart from their wired counterparts, wireless networks tend to have higher mobility, heterogeneity, and higher dependence on the communication medium. It is, thus, important to develop robust wireless systems that can work under a wide range of circumstances. According to the Global Information Technology Report published by the World Economic Forum in 2016~\cite{baller2016global}, the median of mobile network coverage of 148 countries is 99\% of the population, indicating that the cellular technology has advanced enough to cover a variety of different conditions such as extreme weather and altitudes. The high demand for mobile networks leads to complex network designs that require extensive test and validation processes. 

As such, one of the main challenges of wireless network design is to understand how the design would perform in real-life scenarios. A naive way to test such a design is to implement and run all the required network components; yet, this is impractical for most of the scenarios considering modern networks can have thousands of devices or applications. Network simulation is one of the common techniques to avoid full-scale deployment of the network system. It models the network behavior by analytically analyzing the network components such as routers and user equipment. Most of the network simulation uses the discrete event simulation technique where the system is represented with state variables that change at discrete points of time. Network simulators such as NS-3~\cite{ns3} have been widely
used within different network technologies such as cellular networks (e.g., LTE), mobile ad hoc networks (MANET). Their major drawback, however, is the fact that modern network systems become more complex to model by use of analytical methods. Further, modern network testing and validation also require the ability to run different third-party applications such as virtual network functions (VNFs) instead of just testing the proposed model. Network emulation, nonetheless, has the same goal with network simulation, yet allows users to run real applications. In a network emulation, users deploy the software intended to be run on the physical devices over the virtualized network devices. This allows users to test their application with neither deploying a full-scale real network nor analyzing the behavioral model of the network. 

Albeit there exist a few well-known wireless network emulators such as Common Open Research Emulator (CORE)~\cite{ahrenholz2008core}, they suffer several limitations. First, most of the wireless emulators lack configuration and deployment automation to run them to test and validate a variety of network types. Instead, they rely on experienced users to write complex configuration scripts to emulate the different type of networks, which makes the test and validation process be vulnerable to misconfigurations. Numerous studies on network configuration demonstrate that human operator error can account for up to 70\% of system failures~\cite{mahajan2002understanding}. Such failures, therefore, can increase the time for test and validate a network design and network configuration.

Second, existing solutions require a huge number of computing
resources that might not be accessible by most of the users. For example, the Anglova tactical scenario, developed by NATO IST-124 Research Task Group, has a scenario that contains 283 highly mobilized network nodes~\cite{suri2018angloval}. The NATO IST-24 has deployed the Anglova scenario on the US Army Research Laboratory’s Dynamically Allocated Virtual Clustering Management System (DAVC) with the Extendable Mobile Ad-hoc Network Emulator (EMANE)~\cite{emane}. They run a virtual machine (VM) for each network node in the Anglova scenario with a VM for the controller node, requiring to capability to deploy 284 VMs. Unfortunately, running such a large number of VMs requires at least a small scale datacenter which is not accessible or feasible for most researchers. 

Finally, most of the wireless emulators focus on emulating the physical layer (L1) and the data link layer (L2). Despite the possibility of running network layer (L3) or application layer (L4+)\footnote{Throughout this paper, we consider the transport layer within the application layer.} programs such as routing protocols (e.g., OLSR~\cite{clausen2003RFC3626}) or network applications (e.g., iPerf~\cite{iperf}) within these emulators, they do not offer an integrated mechanism to deploy a fully-functional system without configuring each component separately. This leads inexperienced users to spend many hours learning each tool to test and validate their network design. Further, most users use simple testing mechanisms such as running routing protocols with predefined configurations or using basic network applications to check network connectivity such as the Ping~\cite{ping} tool.  Consequently, such users do not have to configure each component independently; rather, it is desirable to have a fully integrated solution that only requires a high-level configuration.

In this work, we present \system, a generic, fully-automated test and validation platform that overcomes the aforementioned limitations of existing emulation systems. First, \system allows users to reuse existing network emulators, consequently, leveraging state-of-the-art solutions. Thanks to recent developments in OS-level virtualization, \system supports different levels of node virtualization, including containerization. This allows for reducing the cost of implementation as well as deployment and test duration. \system, furthermore, supports a variety of network applications and a range of control planes that can be configured via a unified, high-level configuration mechanism. This, subsequently, allows users to configure \system baring only the intended network behavior in their minds instead of configuring each component separately.  


\system contains several layers to achieve generality and full-scale automation. First, it has a base network emulator component that emulates the L1 and L2 behaviors of intended wireless technology. 
Users do not need to configure the base network emulator; instead, they configure \system that automatically configures the base network emulator. Second, \system supports a variety of routing protocols including OSPF~\cite{ospf}, OLSR, OLSRv2~\cite{clausen2014rfc7181}, and OpenFlow-based~\cite{openflow1-4} Software-Defined Networking (SDN). Moreover, users can run multiple protocols on a single node. We specifically analyze such a use case in Section~\ref{ssec:usecases}. As we run each network node on a fully virtualized Linux kernel, users can run any application that can run on a Linux-based system. We also provide a set of network applications including iPerf and Multi-Generator (MGEN)~\cite{mgen} that can generate TCP, UDP, and ICMP traffic.  These applications then analyze the network behavior in detail including latency, throughput, and jitter.

Next, we provide an open-source implementation~\cite{gitemanedocker} of the system that has EMANE as its base network emulator, Docker~\cite{docker} as its virtualization layer, Quagga~\cite{quagga} routing suite with support of OLSR, and OLSRv2 as routing stack, and contains state-of-the-art network performance measurement tools including iPerf, MGEN. Our implementation further runs a resource monitoring stack based on the Telegraf~\cite{telegraf} server agent, Prometheus~\cite{prometheus} time-series database, and Grafana~\cite{grafana} visualization software (TPG stack). The implementation contains 5K+ lines of Python 3 code where users only need to configure a single YAML~\cite{yaml} file and they can use any Linux machine with Docker support. Our implementation, moreover, allows users to run on multiple machines to support thousands of nodes as a Docker Swarm cluster. 

Our extensive experiments show that \system reduces the runtime of emulations by 4x compared to state-of-the-art emulation solutions and reduces the required number of servers by at least 2x up to 5x  compared to state-of-the-art virtualization solutions. Our implementation of \system, further, supports a wide range of routing protocols that including OLSR, OLSRv2, OSPF, RIP, BGP, and SDN, which are not supported by existing wireless network emulators. We further demonstrate that our solution decreases capital expenses (CAPEX) and operational expenses (OPEX) of network emulation over an order of magnitude and reduces the possibility of human configuration errors that causing longer test and validation times. Finally, we discuss a use case of our platform in one of our early work focuses on multiple control plane composition with possible future work.

\section{\system Abstraction}\label{sec:design}

\begin{figure}
    \centering
    \includegraphics[width=\columnwidth]{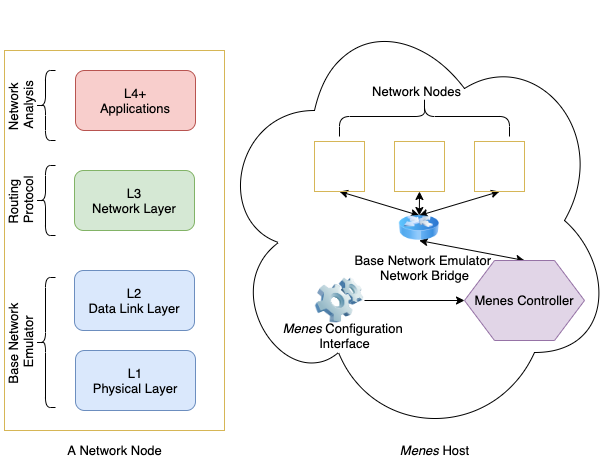}
    \caption{The architecture of \system. (a) A  node has numerous layers responsible for different network layers. (b). A \system host contains several virtual nodes connected via a network bridge. \system is configured via its configuration interface and \system Controller manages each node and the network bridge.}
    \label{fig:menes-model}
    \vspace{-1em}
\end{figure}

In this section, we present an abstract model of \system. We first describe our network model and then give details about different layers of \system. We finally discuss how \system achieves generality and full-scale automation with a single configuration interface.

\subsection{Network Virtualization}
We first discuss how we model a network within our \system. Formally, we define a network as follows:

\begin{definition}[Network]
A network $N$ is a weighted directed graph without multiple edges, where a vertex $n$ is called a \textbf{node} an edge $l$ is called a \textbf{link}.
\end{definition}

Within a network, each node contains a set of functionalities defined by different network layers discussed in the following section. Further, the network links are defined within the base emulation platform. \system creates a virtual connection between nodes using the link specifications given by the user. \system allows users to define heterogeneous links that are different nodes might use different types of physical or data link layers. Moreover, the network can contain a variety of routing protocols and network applications where each node can instantiate a subset of these protocols or applications.  For example, a 10-node network might contain both IEEE 802.11 and 30 MHz very high frequency (VHF) radio, where IEEE 802.11 does not run any routing protocol while 30 MHz VHF radio runs OSPF.

\subsection{Node Virtualization}

\system enables users to run a virtualized version of their intended network devices. Although our implementation in Section~\ref{sec:implementation} focuses on container-based virtualization and Linux-based OSes for generic network devices, our design does not enforce any virtualization technology or OS. \system manages each virtualized node via a network bridge configured with the base network emulator. 

\subsection{L1 and L2: Base Network Emulator}
Apart from other network emulation solutions, \system can run different network emulations within itself thanks to its modular design shown in Figure~\ref{fig:menes-model}. This empowers us to leverage existing work on different wireless technologies, rather than implementing an all-in-one L1 and L2 emulation. We call the L1-L2 emulation within \system as the base network emulator. Combining with the node virtualization technology, \system connects and manages each virtual node via a network bridge.

\subsection{L3: Routing}
\system abstraction allows running any routing protocol that can produce forwarding rules for either IPv4 or IPv6. The dataplane model requires routing protocol to specify (a) Destination (i.e., IP address), (b) Gateway (i.e., next-hop IP address), and (c) Out port (i.e., the interface which connects the node with the specified gateway).

\subsection{L4+: Applications}
Although we focus on network performance measurement and network resource control applications in our implementation in Section~\ref{sec:implementation}, \system abstraction allows users to run any network application that is capable of producing IP packets. In this aspect, we count the transport layers such as TCP and UDP as network applications. For example, our implementation uses the MGEN~\cite{mgen} tool to generate TCP and UDP traffic patterns.

\subsection{\system Controller and Unified Configuration}\label{ssec:configuration}

\begin{figure}
\ttfamily
\begin{tabular}{r L  l}
    (node, link) & \in  & Network $N$
    \\ int-dist(x) & := & uniform(x) | exponential(x) | 
        \\  & & normal(x) | interval(x) | poisson(x)
    \\ network-app & := & mgen | ping | iperf
    \\ emulation &:= & emane | linux | ovs
    \\ routing & := & olsr | olsrv2 | ospf | bgp | rip
    \\ duration & := & \textsc{integer}
    \\ traffic & := & (link, 
        \\  & &(node, node, int-dist)) | \empty
    \\ link & := & (link, 
        \\ & & (node, node, int-dist)) | \empty
    \\ structure & := & ring | full-mesh | random | predefined
    \\ num-nodes & := & int-dist(x)
    \\ topology & := & (num-nodes, structure, link)
    
\end{tabular}
	\caption{A simplified configuration grammar of \system for a given network $N$ }\label{fig:grammar}.

\vspace{-2em}
\end{figure}

One of the major benefits of our system is its unified configuration system that allows users to have an abstract view of different network layers and their complex configuration mechanisms. Instead of individually configuring components of a network, users can define a high-level configuration. To this end, we define a grammar to specify high-level configurations shown in Figure~\ref{fig:grammar}. Our grammar contains base definitions for a network including nodes, topology type, link, and traffic patterns. It is can be extended with the configuration of third-party software; where users can also specify component-specific configuration within the \system's configuration grammar.

\section{Implementation}\label{sec:implementation}

\begin{figure}
    \centering
    \includegraphics[width=\columnwidth]{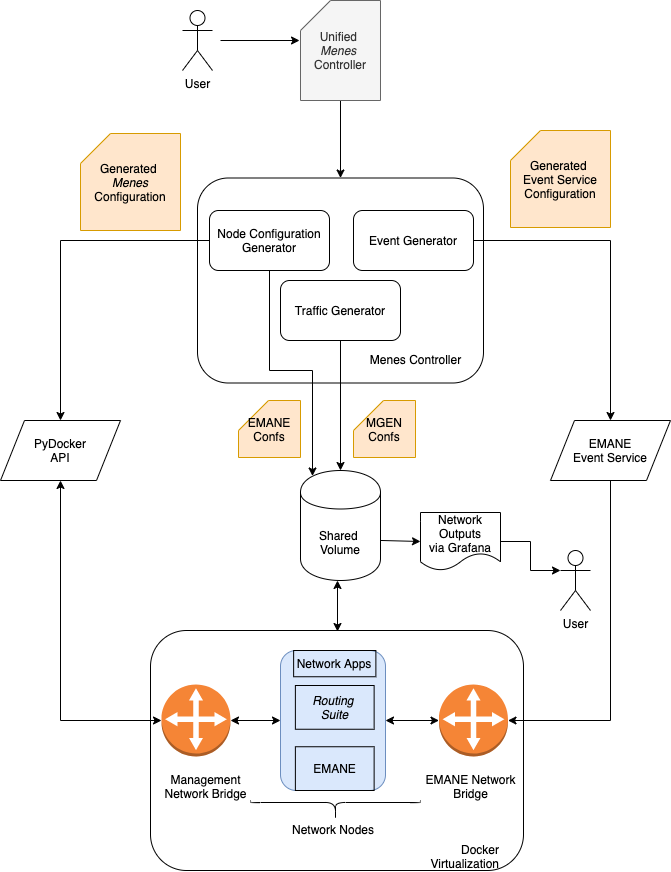}
    \caption{\system implementation workflow. Each node runs the L1-L4 network stack in a Docker container. The nodes are connected via two Docker Network Bridges: (a) EMANE Bridge to emulate wireless connection (b) Management Bridge: To deploy and monitor the containers from \system Controller. The \system Controller configured by the users automatically generate configurations for each application. The users finally access the network performance results and monitoring statistics via Grafana.}
    \label{fig:platform-workflow}
\end{figure}

This section describes the implementation of \system based on its abstraction described in Section~\ref{sec:design}. Our implementation uses Docker containers for the virtualization of nodes in the network and EMANE as the base network emulator. In Section~\ref{ssec:porting}, we discuss a wired implementation of our system where we use Docker Networking as our base emulator. We have provided an open-source
alpha version of this implementation on GitHub \cite{gitemanedocker} and created documentation available at \cite{docsemanedocker}.

\subsection{Node Virtualization with Docker}

\subsubsection{Building a Docker Image}

\begin{figure}[!ht]
    \centering
    \includegraphics[width=0.80\columnwidth]{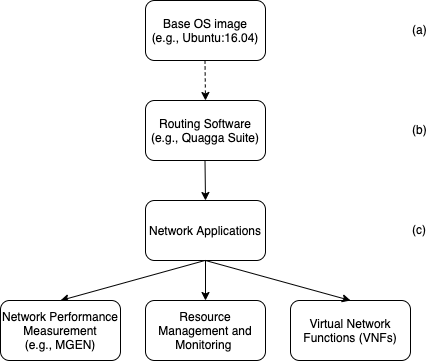}
    \caption{The hierarchy structure of Docker images used in the implementation of \system.  (a) The base OS image such as Ubuntu 16.04. (b) Routing software including OLSR, OLSRv2, and OSPF. (c) Network applications to measure performance and monitor resource usage.}
    \label{fig:docker-hierarchy}
    \vspace{-2em}
\end{figure}

Docker containers run OS-level images that are built automatically by reading \textit{Dockerfile} which is a particular text document containing image building instructions. We model the Docker containers in a similar hierarchical way to the \system abstraction in Section~\ref{sec:design} rather than using a monolithic \textit{Dockerfile} as shown in Figure~\ref{fig:docker-hierarchy}. First, we define a \textit{Dockerfile} to run EMANE individually based on OS image such as Ubuntu 16.04. The \textit{Dockerfile}s in the second layer contain routing software, and we currently support OLSR, OLSRv2, and Quagga routing suite which provides implementations for BGP, OSPF, RIP. This layer can be also merged with a \textit{Dockerfile} containing OpenVSwitch to run OpenFlow-based SDN control planes. The final layer contains numerous network applications, including iPerf, MGEN, and our TPG monitoring stack. Preserving this hierarchy, users can also change their base network emulator, routing protocol, or network applications.

\subsubsection{Docker on a Single Machine}
\system can run on a single host machine that can run Docker and EMANE. Figure~\ref{fig:platform-workflow} shows the architecture of our implementation on a host machine, where each node is represented as a Docker container and connected via Docker Network Bridge. Each container runs a set of applications, Telegraf as monitoring software, and EMANE instance automatically configured via \system Controller. The Docker Network Bridge itself also connected to a single EMANE process that generates the link events based on the user configurations. The Management Network Bridge, on the other hand, is a 1-hop network where each node is directly connected to the \system Controller enabling us to configure nodes as well as reading the metrics from Telegraf via Prometheus. The Management Network Bridge does not have any effect on the EMANE Network Bridge that carries the experimentation traffic.
\subsubsection{Clustering with Docker Swarm}
Despite the recent advancements on multicore processors and multiprocessor servers, there is a practical limit on the maximum number of concurrent containers that can be run on a single server. We, therefore, extend \system to support running on multiple servers to enable users to emulate large topologies. To do so, we run Docker on swarm mode where a server is selected as a controller that runs as a Docker manager and also runs the \system controller. We then run the other servers as Docker workers that only run containers to emulate network nodes. This mode only requires users to configure their servers to accessible via SSH from the controller node and configure \system with the management IP addresses of each server.

\begin{figure}
    \centering
    \includegraphics[width=0.80\columnwidth]{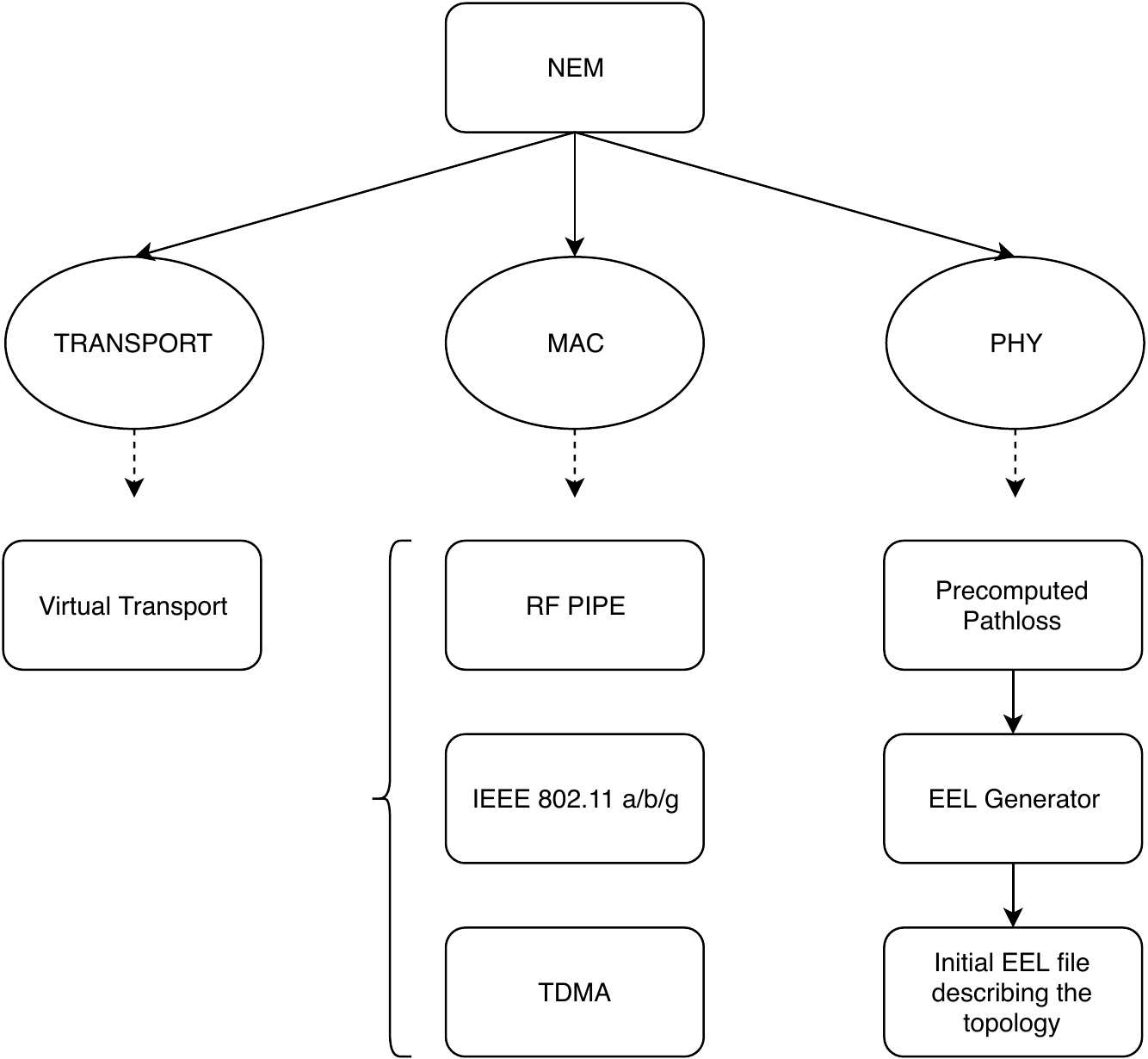}
    \caption{EMANE configuration structure. A NEM configuration describes three layers: The Transport, MAC, and PHY layers. Transport and PHY layer supports a single type of instantiation where the MAC layer can be a type RF PIPE, IEEE 802.11 a/b/g, or TDMA. }
    \label{fig:emane-config}
    \vspace{-1em}
\end{figure}

\subsection{Network Emulation with EMANE}

As we discuss in the earlier sections, each container has an interface that is connected to the EMANE Network Bridge in the host. Each EMANE process runs a set of Network Emulation Modules (NEMs) that are network stacks as interfaces. Figure~\ref{fig:emane-config} shows the different layers in a NEM and possible configuration options for each layer provided by EMANE.

\subsubsection{L1: Physical Layer Emulation}
The first network layer emulated with EMANE is the physical layer. EMANE supports pre-computed path loss files or dynamic event generation via EMANE Emulation Event Log (EEL) Generator. \system uses EEL Generator and it extends this into two options. First, users can pick a pattern of events such as Poisson arrival events. Second, they can also give a precomputed events as such in the NATO IST-124 Anglova scenario. The \system Controller reads these user configurations and feeds into the EMANE EEL Generator running in the host machine, which then triggers the events on particular nodes via EMANE Network Bridge.

\subsubsection{L2: MAC Emulation}
The MAC layer contains different radio models including a simple pipe model (RF Pipe), IEEE 802.11 a/b/g, TDMA, and LTE. In \system implementation, we instantiate the first three radio models per our experiments focus on MANETs. Each model has its own configuration parameters that are managed automatically via the \system Controller. For instance, if a user configures \system to have a network at rate 54 Mbps, it will automatically translate this into 802.11g with OFDM. However, the user can also give more specific instructions to pick the exact MAC layer he or she wants to run like RF Pipe.

\subsubsection{Virtual Transport Layer}
In this layer, we configure EMANE to run a virtual transport that is capable of both IPv4 and IPv6 addresses and packet processing. Users, therefore, can deploy any version of routing protocols (e.g., OSPFv4 or OSPFv6). NEMs in this mode have unicast, broadcast, or multicast capabilities; that enable users to have a complete network device emulation.

\subsection{L3: Running Routing Protocols}
First, we investigate the control plane at the network layer. Our \system implementation has numerous routing protocols including wireless oriented OLSR and OLSRv2 as well as traditional routing protocols such as OSPF and BGP provided by the Quagga Routing Suite. The users should only specify intended routing protocols to run with an option of different protocols for a different subset of nodes, the \system controller then configures the specified protocols automatically. Further, the users can also give particular configurations for routing software with the \system configuration grammar defined in Section~\ref{ssec:configuration}. 

The forwarding plane, on the other hand, is the Linux routing stack by default. Yet, the users can also specify to run specific forwarding planes including virtual switches such as OpenvSwitch~\cite{ovs}. This, therefore, allows users to emulate not only traditional routing protocols but also modern SDN applications. In our implementation, we test this capability with a simple OpenFlow-based SDN controller.

\subsection{L4+: Network Performance Measurement}

In our \system implementation, we mainly focus on network performance measurement applications as they are fundamental to test and validate a network deployment. To accomplish this goal, we use three state-of-the-art network performance measurement applications that can generate TCP, UDP, or ICMP traffic loads: Ping, iPerf, and MGEN.

We use Ping for simplistic network connectivity validation, where we use iPerf to generate TCP or UDP loads to test not only the connectivity but also throughput and jitters. MGEN, developed by the U.S. Naval Research Laboratory,  is a sophisticated toolset that can generate real-time traffic patterns for both TCP and UDP applications.

\subsection{Resource Management and Monitoring}
One challenge in wireless networks is the fact that most devices have very limited resources. For example, a handheld device runs on a small lithium-ion battery that can only last for a couple of hours in communication. It is, therefore, crucial for network designers to understand the resource consumption of the applications they are running on the devices. To this end, we developed an automated resource control mechanism running each network node device and reporting back to the controller. 

Our implementation is based on two state-of-the-art data collection and storage agent. We run Telegraf to collect and send metrics from network nodes to the \system Controller which runs the Prometheus time-series database to collect and analyze the metrics. We further run Graphana with built-in charts and graphs for users to visualize the metrics stored in Prometheus. 


\subsection{Continuous Integration and Continuous Deployment (CI/CD)}
One aspect of network emulation has an equivalent version of the real network and test each change in the network by the emulation first. To meet this goal, we integrate our implementation of \system with Travis CI/CD service. Users can deploy Docker Swarm over public clouds such as Google Cloud, then, changing \system configuration triggers auto build and deploy of the emulation with new configurations.
\section{Experiments and Evaluation}

In this section, we demonstrate the benefits and use cases of our platform. 

Our extensive experiments show that \system reduces the runtime of emulations by 4x compared to state-of-the-art emulation solutions and can support a wide range of routing protocols that including OLSR, OLSRv2, OSPF, RIP, BGP, and SDN. We further demonstrate that our solution decreases CAPEX and OPEX costs of network emulation over an order of magnitude and reduces the possibility of human configuration errors that causing longer test and validation duration.

\begin{table}[]
    \centering
    \begin{tabular}{|c|c|c|c|}
        \hline 
         ~ & DAVC & Docker Swarm  \\ \hline 
         Average Bootstrap Time & 123 s & 29 s\\ \hline
    \end{tabular}
    \caption{Average bootstrap time for a subset of NATO IST-124 Anglova scenario with 30 nodes.}
    \label{tab:avgdown}
    \vspace{-2em}
\end{table}

\subsection{Runtime Evaluation}

The table shows the bootstrap time of a subset NATO IST-124 Anglova scenario with 30 nodes using the DAVC and Docker Swarm cluster on Google Cloud. As we see in the table, Docker Swarm can reduce the initialization time by 4x. This is because starting containers does not require to bootstrap the Linux kernel, which is required to start the VMs.

\subsection{Overhead Scaling}\label{ssec:overhead}

\begin{figure}[!ht]
    \centering
    \includegraphics[width=\columnwidth]{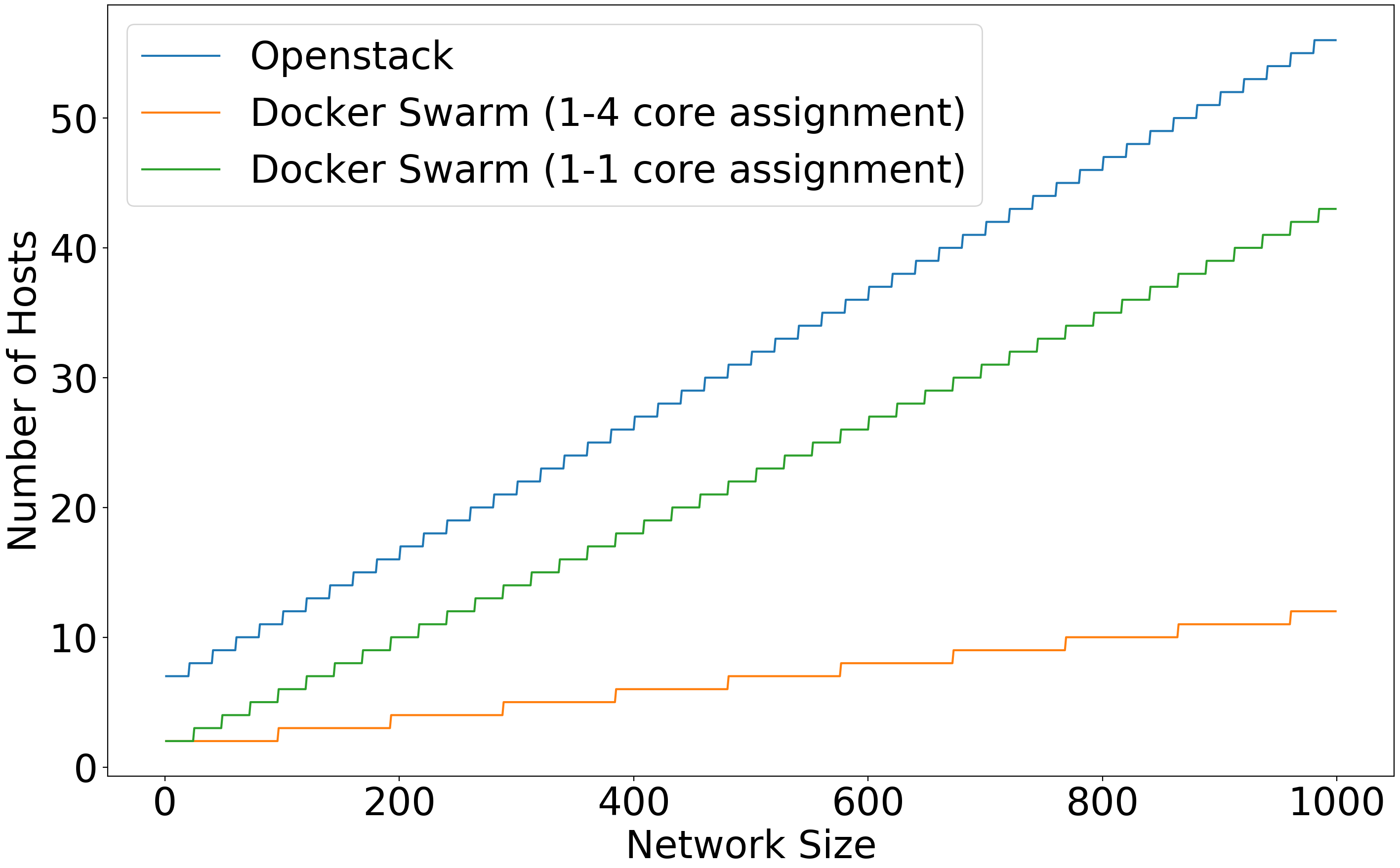}
    \caption{The number of hosts required to emulate full mesh networks with different sizes, where each host has Intel Xeon E5 @2.30GHz with 24 cores. (a) OpenStack~\cite{openstack} private cloud with 3 controller and 3 storage hosts. (b) Docker Swarm that assigns each CPU core to a container. (c) Docker Swarm allows each node to have up to 88 containers.}
    \label{fig:result-number-of-hosts}
\end{figure}

In this section, we demonstrate how our \system implementation scales with the network size. Figure~\ref{fig:result-number-of-hosts} shows three different scenarios with OpenStack and Docker virtualizations. In each scenario, we calculate the number of hosts required to run to emulate a given network size. Our results show that using Docker Swarm can reduce the number of required hosts by 2x to up to 5x.

\subsection{Cost Efficiency}\label{ssec:cost}

\begin{figure}[!ht]
    \centering
    \includegraphics[width=\columnwidth]{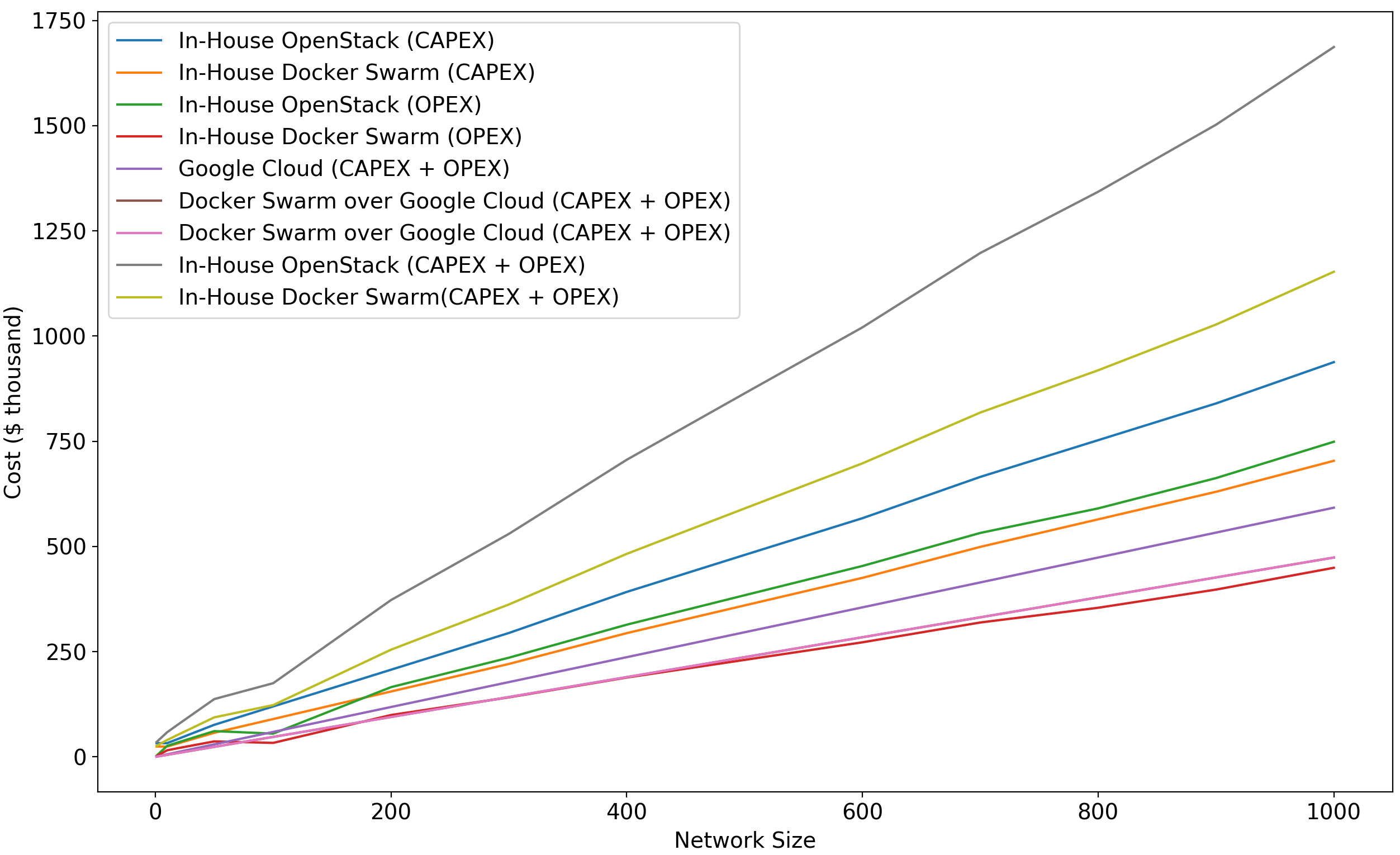}
    \caption{Average cost of running dedicated network emulation for two-years. The figure compares (a) In-House OpenStack and Docker Swarm deployments and (b) Google Cloud VM and Docker Swarm deployments.}
    \label{fig:result-network-cost}
\end{figure}

The second set of evaluation demonstrates the benefits of \system considering CAPEX and OPEX of network emulation. We used TCO tool~\cite{tco} to compute the total cost of running network emulation over a year, where we consider the cost in-house servers as CAPEX and cloud expenses and management expenses as OPEX. We consider two different deployment environments for our scenarios. First, we consider the case that a user deploys OpenStack private cloud where each server has 24 CPU cores. Second, we investigate a public cloud service provided Google Cloud~\cite{gcloud}, where each VM runs at the minimum possible resources (i.e., 1 vCPU with 2GB RAM at an hourly rate of \$24.67). We then test the cost of the same scenarios that we described in Section~\ref{ssec:overhead} as shown in Figure~\ref{fig:result-network-cost}.

\subsection{Use cases}\label{ssec:usecases}
As \system emerged as a part of our existing project in~\cite{li2019magnalium}, it has been tested in numerous use cases. We mostly focused on MANET use cases for military including the emulation of the NATO IST-124 Anglova~\cite{suri2018angloval} scenario. We further deploy multiple control planes within \system namely OLSR, OLSRv2, and OSPF with a centralized OpenFlow-based SDN controller. The experiments include use cases such as  IGP Migration, fast-failure recovery, network verification; as well as, network stress testing where we run a couple of hundred control planes at the same time on the state-of-the-art network topologies such as Stanford backbone network~\cite{stanfordnetwork}. We further use an implementation of \system for wired networks that uses OpenvSwitch and Linux networking stack as its base network emulator. 

\subsection{Extensibility}

A major promise of \system is 
high extensibility thanks to its modular design. Although our implementation is limited to a set of software, the users can extend \system according to their needs. Each component can be replaced with a counterpart. For instance, if a user wants to test a different routing protocol that our implementation does not have, simply modifying the L3 \textit{Dockerfile} and adding auto-configuration scripts to the \system Controller would be sufficient. We further discuss such an extension for wired networks in Section~\ref{ssec:porting}.
\vspace{-1em}
\section{Related Work}

\para{Network Emulators.} There have been numerous tools developed for network emulation namely Common Open Research Emulator (CORE)~\cite{core}, EMANE~\cite{emane}, and NetSim~\cite{rathi1990new}. The CORE is a tool for emulating networks on multiple machines that can be connected to live networks. It contains a graphical interface to manage topologies, and Python bindings to script the network emulation. EMANE, on the other hand, is a framework for real-time modeling of mobile network systems that focuses on physical layer wireless emulation. Network Emulation Modules (NEMs) in EMANE can be heterogeneously used with real and virtualized network stacks. It further provides an event-driven control bus and logging facilities. NetSim is a full-stack packet-level network simulator and emulator that has been used by civil and military use cases over a couple of decades. NetSim allows users to integrate new protocols or network devices by requiring lower cost and in less time compared tho hardware prototypes.

\para{Virtualization.} Kernel-based Virtual Machine (KVM)~\cite{kvm} is the state-of-the-art virtualization module within the Linux kernel allowing the kernel to run as a hypervisor of guest virtual machines (VMs). QEMU~\cite{qemu} is a VM emulator that emulates the machine processor via binary translation of instructions, which provides a set of hardware and device models. QEMU and KVM can be used together in Linux host machines which can be integrated with hardware acceleration to achieve higher performance. OS-level virtualization, on the other hand, is the technology that allows the existence of multiple isolated user spaces sharing the same kernel with the host~\cite{osvirtualization}. This technique is different from running VMs where each VM instantiates its own kernel. Linux Containers (LXC)~\cite{lxc} is a container platform integrated to the Linux kernel providing user space for simple applications to run over containers. Docker~\cite{docker} is a set of the platform as a service product that also has OS-level virtualization to deliver software in packages in containers. It offers better isolation compared to LXC, and it further presents complex network capabilities. Kubernetes~\cite{kubernetes} is another system for automating deployment, scaling, and management of containerized applications that mostly focuses on micro-services. Further, major public cloud vendors such as Amazon Web Services (AWS)~\cite{aws} offers application-specific virtualization such as serverless computing.

\section{Conclusion}

This paper proposes \system a generic, fully-automated test and validation platform for wireless networks. \system allows users to easily deploy and iterate homogeneous or heterogeneous networks using different physical layer technologies or routing suites. It further allows users to deploy real applications or Linux-based VNFs. It is one of the first wireless network emulators enabling users to deploy thousands of virtual network devices within minutes by use of less computing
resources compared to traditional solutions. 
\section{Recommendations}
This work introduces the idea of full-stack emulation of wireless networks that enables users to use high-level configuration mechanisms rather than configuring individual components separately. We give a base implementation of \system  that focuses on MANET use cases; yet, it lacks support for other wireless technologies such as cellular networks. Further, we only focus on network performance and resource monitoring in our implementation, and a further direction can implement the state-of-the-art VNFs within our emulation platform. Thanks to its modular design, the developers only need to add components and define how high-level configuration grammar would parse into the particular configuration of their component. In the following section, we discuss our experience in porting \system implementation into wired networks.

\subsection{Porting to Wired Networks}\label{ssec:porting}
Although our work is focused on wireless networks, one aspect of our design is the modularity of the core network emulation component. As we discussed in Section~\ref{ssec:usecases}, the core network emulation component can be replaced with a wired network emulator. As wireless networks are highly mobilized and more complex, we pick wireless emulation as the starting point so that our platform can easily work with wired networks requiring only to change its base network emulator with a wired network emulator such as Linux virtual networking stack.
\begin{acks}

The author would like to thank the Yale research team lead by Prof. Y. Richard Yang and his colleagues Dr. Geng Li, Dr. Qiao Xiang, Shenshen Chen, and Dong Guo for their support and assistance with this project and its related publications. The author would also like to thank the coauthors of ~\cite{li2019magnalium}: Akrit Mudvari from Yale University, Patrick Baker and Jeremy Tucker from UK Defence Science and Technology Laboratory, Sastry Kompella from U.S. Naval Research Laboratory,  Kelvin M. Marcus and Paul Yu from U.S. Army Research Laboratory, and Franck Le from IBM T.J. Watson Research Center for their open comments and discussions. Finally, the author would like to thank Emir Alaattin Yilmaz from Sabanci University for his comments that immeasurably improved the manuscript.


This research was sponsored by the U.S. Army Research Laboratory and the U.K. Ministry of Defence under Agreement Number W911NF-16-3-0001. The views and conclusions contained in this document are those of the authors and should not be interpreted as representing the official policies, either expressed or implied, of the U.S. Army Research Laboratory, the U.S. Government, the U.K. Ministry of Defence or the U.K. Government. The U.S. and U.K. Governments are authorized to reproduce and distribute reprints for Government purposes notwithstanding any copyright notation hereon.
\end{acks}

\bibliographystyle{ACM-Reference-Format}
\bibliography{main}


\end{document}